# Predicting Coastal Water Levels in the Context of Climate Change Using Kolmogorov-Zurbenko Time Series Analysis Methods

**Barry Loneck***, **Igor Zurbenko, and Edward Valachovic**

**Department of Epidemiology and Biostatistics**
**School of Public Health**
**The University at Albany**
**State University of New York**

## ABSTRACT

Given recent increases in ocean water levels brought on by climate change, this investigation decomposed changes in coastal water levels into its fundamental components to predict maximum water levels for a given coastal location. The study focused on Virginia Key, Florida, in the United States, located near the coast of Miami. Hourly mean lower low water (MLLW) levels were obtained from the National Data Buoy Center from January 28, 1994, through December 31, 2023. In the temporal dimension, Kolmogorov-Zurbenko filters were used to extract long-term trends, annual and daily tides, and higher frequency harmonics, while in the spectral dimension, Kolmogorov-Zurbenko periodograms with DiRienzo-Zurbenko algorithm smoothing were used to confirm known tidal frequencies and periods. A linear model predicted that the long-term trend in water level will rise 2.02 feet from January 1994 to December 2050, while a quadratic model predicted a rise of 5.91 during the same period. In addition, the combined crests of annual tides, daily tides, and higher frequency harmonics increase water levels up to 2.16 feet, yielding a combined total of 4.18 feet as a lower bound and a combined total of 8.09 feet as an upper bound. These findings provide a foundation for more accurate prediction of coastal flooding during severe weather events and provide an impetus for policy choices with respect to residential communities, businesses, and wildlife habitats. Further, using Kolmogorov-Zurbenko analytic methods to study coastal sites throughout the world could draw a more comprehensive picture of the impact climate change is having on coastal waters globally.

* Corresponding Author Email: bloneck@albany.edu

# Predicting Coastal Water Levels in the Context of Climate Change Using Kolmogorov-Zurbenko Time Series Analysis Methods

In light of recent increases in ocean water levels brought on by climate change, this investigation estimates the impact of long-term trends in concert with known tides and their higher frequency harmonics on change in coastal ocean water levels for a major Southeastern city in the United States. To do this, change in total water level was deconstructed into its fundamental components then, using parameter estimates, predicted future minimum and maximum coastal water levels. Such predictions are needed to prepare for severe weather events such as hurricanes and tropical storms, in the short run, and to prepare for shrinking coastlines of major cities and wildlife habitats, in the long run.

## 1. Background

Climate change is leading to a significant rise in ocean levels throughout the world and is the result of two primary factors (Zurbenko et al., 2020; Zurbenko & Potrzeba-Macrina, 2019b). The first is solar irradiation caused by sunspot activity and it serves to increase atmospheric temperatures (Potrzeba-Macrina & Zurbenko, 2019). The second is atmospheric pollution caused by human activity, which traps solar irradiation, thus adding to increases in atmospheric temperatures (Potrzeba-Macrina & Zurbenko, 2020; Zurbenko & Potrzeba-Macrina, 2019a). These increases in atmospheric temperatures have led to the melting of glaciers and the polar ice caps, thereby increasing ocean levels. However, rising ocean levels lead to increases in atmospheric water vapor (Potrzeba-Macrina & Zurbenko, 2021; Zurbenko & Potrzeba-Macrina, 2021; Zurbenko & Sun, 2015, 2016) which traps even higher levels of solar irradiation. Consequently, rising ocean levels constitute a secondary cause of climate change.

Tides also directly affect coastal water levels. In recent history, Darwin (1997) and Doodson and Lamb (1997) systematically investigated and delineated the periods and frequencies in annual tides, caused by the earth's revolution around the sun, as well as periods and frequencies in daily tides, caused by the earth's rotation under the moon and the sun (Cartwright, 2000; Kowalik & Luick, 2019; Webb, 2021). In addition, higher frequency harmonics of annual and daily tides also affect coastal water levels; indeed, while Doodson initially identified 388 such harmonics, several thousand partial waves are now known to exist (Casotto & Biscani, 2004).

Rising ocean levels in concert with preexisting annual and daily tides, as well as their higher frequency harmonics, can be particularly problematic for coastal towns and cities, threatening residential communities, businesses, and wildlife habitats, and stand to get worse over time. Using Kolmogorov-Zurbenko time series analysis methods, this investigation examined changes in ocean water levels by deconstructing it into its fundamental components: (1) changes due to long-term trends attributable to climate change; (2) changes due to annual tides attributable to the earth's elliptical orbit around the sun; (3) changes due to daily tides attributable to the earth's rotation relative to both the sun and moon; and (4) changes in water level height attributable to the higher frequency harmonics of annual and daily tides.

To conduct this analysis, two state-of-the-art time series analysis methods were used. In the temporal dimension, Kolmogorov-Zurbenko (KZ) filters (Zurbenko et al., 2020) separated raw time series data into its long-term trend, annual tides, daily tides, and harmonic components. KZ filters were used because they provide high resolution of signal (i.e., tidal) frequencies (Rao et al., 1997; Zurbenko, 1986) and because they are robust in the face of missing data (Iasonos & Zurbenko, 2002). In the spectral dimension, Kolmogorov-Zurbenko (KZ) periodograms with DiRienzo-Zurbenko algorithm (DZA) smoothing (Zurbenko et al., 2020) confirmed primary annual and daily tidal frequencies and periods. The KZ periodogram is a log-periodogram that uses the Kolmogorov-Zurbenko Fourier transform (i.e., an iterated Fourier transform) to estimate signal frequencies (Yang & Zurbenko, 2010; Zurbenko, 1986), while DZA smoothing (DiRienzo & Zurbenko, 1999) ensures a high degree of sensitivity, accuracy, and resolution in detecting, identifying, and separating those signal frequencies, and is robust with high levels of missing data (Loneck et al., 2024; Potrzeba-Macrina & Zurbenko, 2017; Zurbenko et al., 2020). The results of the time series analyses served as the foundation for a linear model and a quadratic model, both of which were subsequently used along with the annual, daily, and harmonic minima and maxima to predict long-term minimum and maximum coastal water levels over time. These predictions can be used to prepare for emergency services prior to the onset of severe weather events, in the short run, and to prepare for the preservation and relocation of coastal communities, businesses, and wildlife habitats, in the long run.

## 2. Methods

To estimate the impact of long-term trends, annual and daily tides, and higher frequency tidal harmonics on changes in coastal ocean water levels, we conducted a time series analysis. This analysis included site selection and identification of a data source, construction of a dataset, and implementation of time series analysis methods comprised of temporal analyses, spectral analyses, and predictions of future coastal water levels.

### 2.1 Site Selection and Data Source

Miami-Dade County, Florida, in the United States was selected as the site for this investigation because its National Risk Index is 99.8, with a Hurricane Risk Index of 100, a Coastal Flooding Index of 76.5, and a Riverine Flooding Index of 99.8 (Federal Emergency Management Administration, n.d.).

Data was obtained from the National Oceanic and Atmospheric Administration's (NOAA) National Data Buoy Center (US Department of Commerce, n.d.) for Virginia Key, Florida (https://www.ndbc.noaa.gov Station VAKF1). Virginia Key is located approximately 2.5 miles southeast of Miami, Florida, and approximately 3.3 miles south-southwest of Miami Beach, Florida.

### 2.2 Dataset

The dataset consists of the verified hourly water levels from January 28, 1994, 16:00 (when station verification became active) to December 31, 2023, 23:00 for a total of 262,304 observations. Water level was measured using the mean lower low water (MLLW) level. The MLLW was measured in feet and is relative to a reference point (0) that is "the average of the lower low water height of each tidal day observed over the National Tidal Datum Epoch [i.e., 1983 through 2001]" (National Oceanic and Atmospheric Administration, n.d.). Use of MLLW is consistent with previous investigations by Zurbenko and Potrzeba-Macrina (2021). Although there are several measures of water level (i.e., mean higher high-water level, mean high water level, mean sea level, mean low water level, mean lower low water level), a key difference among them is the reference point from which deviations are measured. Thus, converting from one measure to another is typically done by adding or subtracting the difference in reference points to the observed water level (Berg, 2016).

There were two intervals of missing data during the period from January 28, 1994, to December 31, 2023. The first was from October 3, 1997, 16:00 to November 8, 1997, 14:00, with a total of 863 missing observations, and the second was from February 14, 2016, 02:00 to February 25, 2016, 14:00, with a total of 277 missing observations. No other data was missing.

### 2.3 Data Analysis Methods

In the temporal dimension, long-term trends, annual tides, daily tides, and higher frequency harmonics were extracted using Kolmogorov-Zurbenko (KZ) filters (Close et al., 2020). The KZ filter is a low pass filter which can also be made to function as a high pass filter to effectively separate and smooth time series data into separate time scales (i.e., long-term trends, annual tides, daily tides, higher frequency harmonics); its function as a high pass filter is accomplished by subtracting a low frequency component from the given time series which yields the remaining higher frequency signals. Thus, one can shuttle back and forth from low pass filtering to high pass filtering to identify time series frequency components (Valachovic, 2014). With the KZ filter, time scale separation is done by specifying the algorithm's moving average window width, $m$, and smoothing is done by specifying the number of iterations, $k$, of the moving average.

Extraction of the long-term trend and annual tides were done using the entire dataset from 1994 through 2023. However, because the KZ filter can only accommodate periods of missing data less than 3 times the window width and because the window width necessary to extract daily tides was set at $m=6$, extraction of daily tides was limited to the period from February 25, 2016, to December 31, 2023, for a total of 68,793 observations.

In the spectral dimension, frequencies within the separate time scales were identified using Kolmogorov-Zurbenko (KZ) periodograms with DiRienzo-Zurbenko algorithm (DZA) smoothing (Close et al., 2020). Unless otherwise specified: the initial window width of the KZ periodogram was set at $m_{Initial} = 1000$ so that frequency estimates could be made with an accuracy of 0.001; the number of iterations was set at $k=3$; and the DZ proportion of smoothing was set at 0.005 to assure adequate detection of tidal frequencies. Each KZ periodogram with DZA smoothing also included computation of 95% confidence intervals for respective signal strength (Loneck et al., 2025) within the respective time scale.

For each periodic component (i.e., annual tides, daily tides) and for higher frequency harmonics, its minimum contribution to water level and its maximum contribution to water level were set equal to their corresponding empirical 2.5 and 97.5 percentiles. In turn, the 2.5 and 97.5 empirical percentiles were computed using the linear interpolation between two observed quantile plotting positions, $p_k = \left(k - \frac{1}{3}\right)/\left(n + \frac{1}{3}\right)$, with $\frac{k}{n}$ the desired percentile and $n =$ the number of observations. This percentile formula was used because it yields median-unbiased estimators which are independent of an underlying distribution (Hyndman & Fan, 1996). To remain consistent across the tidal components and higher frequency harmonics, medians, minima (2.5 percentile), and maxima (97.5 percentile) were computed for the limited period of February 25, 2016, to December 31, 2023.

To predict future water levels, two regression models of the long-term component were used in concert with the respective minima and maxima of annual tides, daily tides, and higher frequency harmonics. The first model was linear and consisted of a simple linear regression of the long-term component on *time*, providing a lower bound for its contribution to predicted water levels. However as noted earlier, climate change has been the result of two primary factors: solar irradiation and the influence of human activity. Solar irradiation is proportional to sunspot activity which has known long-term periodicities and has been in a long-term trough in the most recent decades, yet ocean water levels continue to rise due to human factors (Potrzeba-Macrina & Zurbenko, 2021). Nonetheless, sunspot activity and, thus, solar irradiation is beginning a long-term upswing in its periodicity (Zurbenko & Potrzeba-Macrina, 2019b, 2019a). Consequently, the second model was quadratic and consisted of regressing the long-term component on *time* and $time^2$ to provide an upper bound of its contribution to predicted water levels.

Each model's regression was done for the period of February 25, 2016, to December 31, 2023, and the respective intercepts and regression coefficients were used to predict long-term water levels from January 1, 2024, to December 31, 2050. Finally, the minima and the maxima for the annual tides, daily tides, and higher frequency harmonics were summed and used to construct minimum and maximum water levels for both models across the entire timeframe, from January 28, 1994, to December 31, 2050, so as to provide greater perspective of changing water levels.

## 3. Results

Time series analyses proceeded in five stages: analysis of the raw time series dataset, analysis of long-term trends, analysis of annual tides, analysis of daily tides, and examination of higher frequency harmonics. In addition, the findings were summarized and used to predict water levels through the year 2050.

### 3.1 Analysis of Raw Time Series Dataset

A plot of the raw time series data and its respective KZ periodogram for water levels at Virginia Key, Florida from January 28, 1994, to December 31, 2023, are presented in Figure 1. Overall, the time series indicates a gradual upward trend in water level. In the KZ periodogram, two sets of primary frequencies are apparent. In the first, the strongest is the solar diurnal tide with a frequency of $\lambda = 0.042$ (period of 23.81 hours), followed by the lunar diurnal tide with a frequency of $\lambda = 0.039$ (period of 25.64 hours) and, finally, the larger lunar elliptical diurnal tide with a frequency of $\lambda = 0.037$ (period of 27.03 hours). In the second set, the strongest is the principal lunar semi-diurnal tide with a frequency of $\lambda = 0.081$ (period of 12.35 hours), followed by the principal solar diurnal tide with a frequency of $\lambda = 0.083$ (period of 12.05 hours). Taken together and given the arguments selected for the KZ periodogram with DZA smoothing, these two sets of frequencies are within the periodogram's limits of accuracy for the diurnal and semi-diurnal tides present along the East Coast of North America. The higher frequencies (with commensurate shorter periods) that can be seen above these two sets in the KZ periodogram are higher frequency harmonics of the two sets of primary frequencies.

### 3.2 Analysis of Long-Term Trends

To identify long-term trends and possible long-term frequencies, a KZ filter with a window of 1 year ($m = 8760$) and 3 iterations ($k = 3$) was used. As shown in Figure 2, the time series indicated the water level increased .76 feet over the 30-year period, providing evidence that ocean levels are rising on the Miami, Florida coast. However, the KZ periodogram did not show evidence of long-term periodicity in the water level data; nevertheless, long-term periodicity greater than 30 years may be present but undetectable because the length of time in this dataset is too short.

### 3.3 Analysis of Annual Tides

The next step was to extract the annual tidal component from the residual of the long-term component. Residuals of the long-term component were determined by subtracting the long-term component from the raw time series data and as expected, the long-term residuals centered around 0. The annual tidal component was extracted from these residuals using a KZ filter with a window of 29 days ($m = 696$) and 3 iterations ($k = 3$). As shown in Figure 3, periodicity was evident in this time series, but exact frequencies could not be discerned in its KZ periodogram because the rate at which data was collected (i.e., 1 observation per hour) was too fine. Consequently, a more detailed analysis was necessary through an examination of tides over the course of a calendar year.

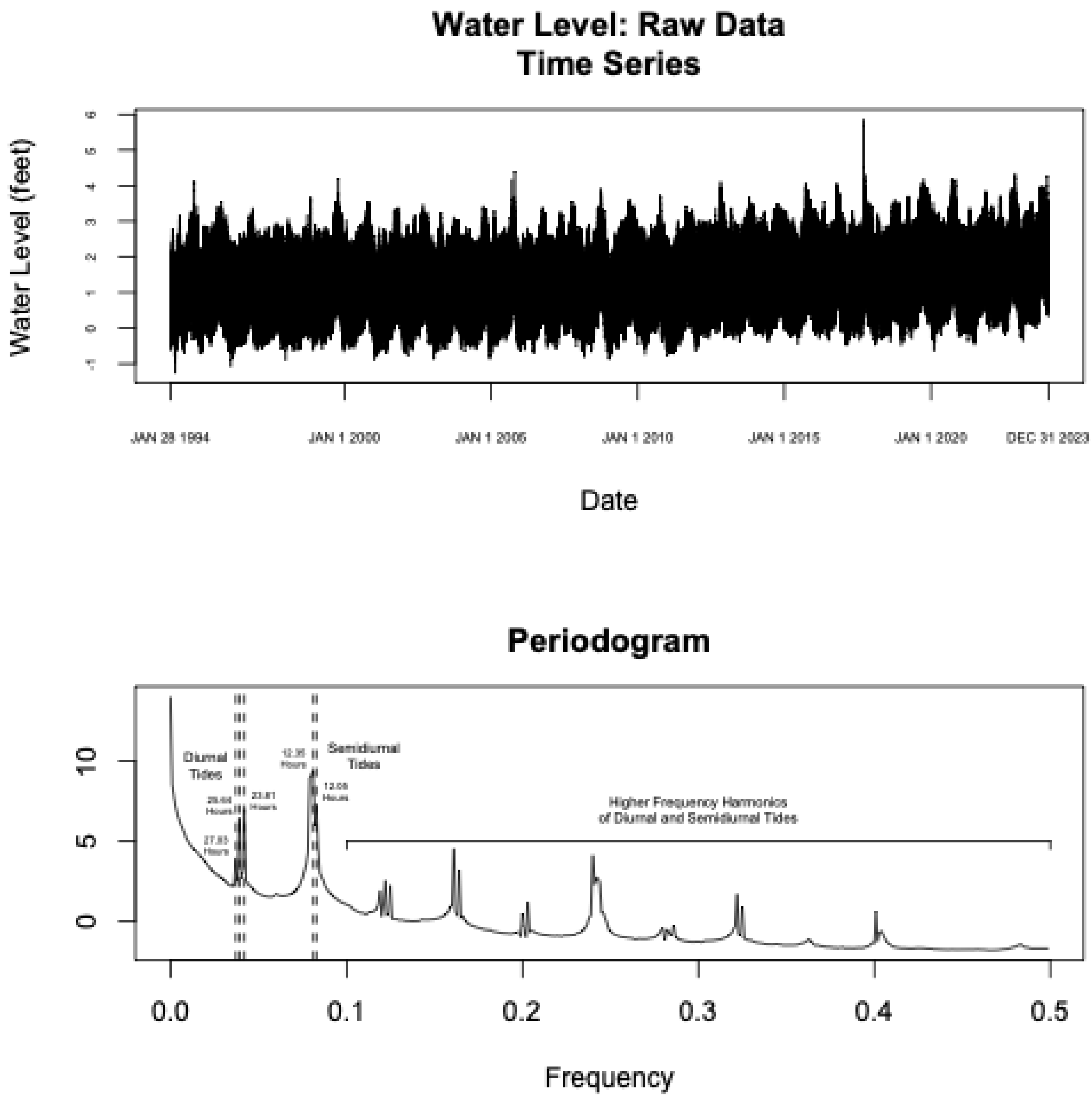

**FIGURE 1. Time Series* and Periodogram** for Hourly Water Levels at Virginia Key, Florida from January 28, 1994, to December 31, 2024.**

***Time series displays a gradual upward trend in water level.**

****Periodogram displays diurnal tides (27.03, 25.6, 23.81 hours) and semidiurnal tides (12.35, 12.05 hours).**

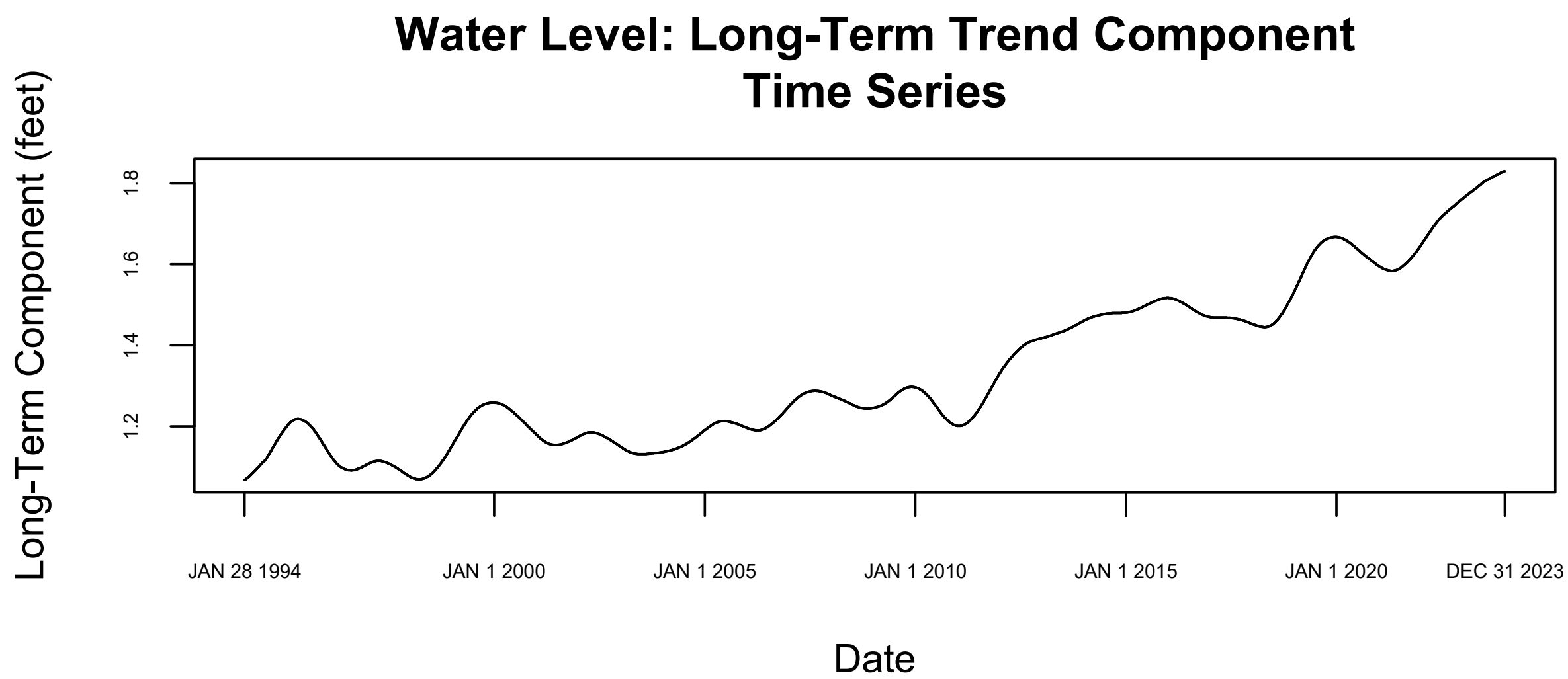

**FIGURE 2. Time Series of the Long-Term Component for Hourly Water Level at Virginia Key, Florida from January 28, 1994, to December 31, 2024.**
*** Window width $m = 8760$; number of iterations $k = 3$**

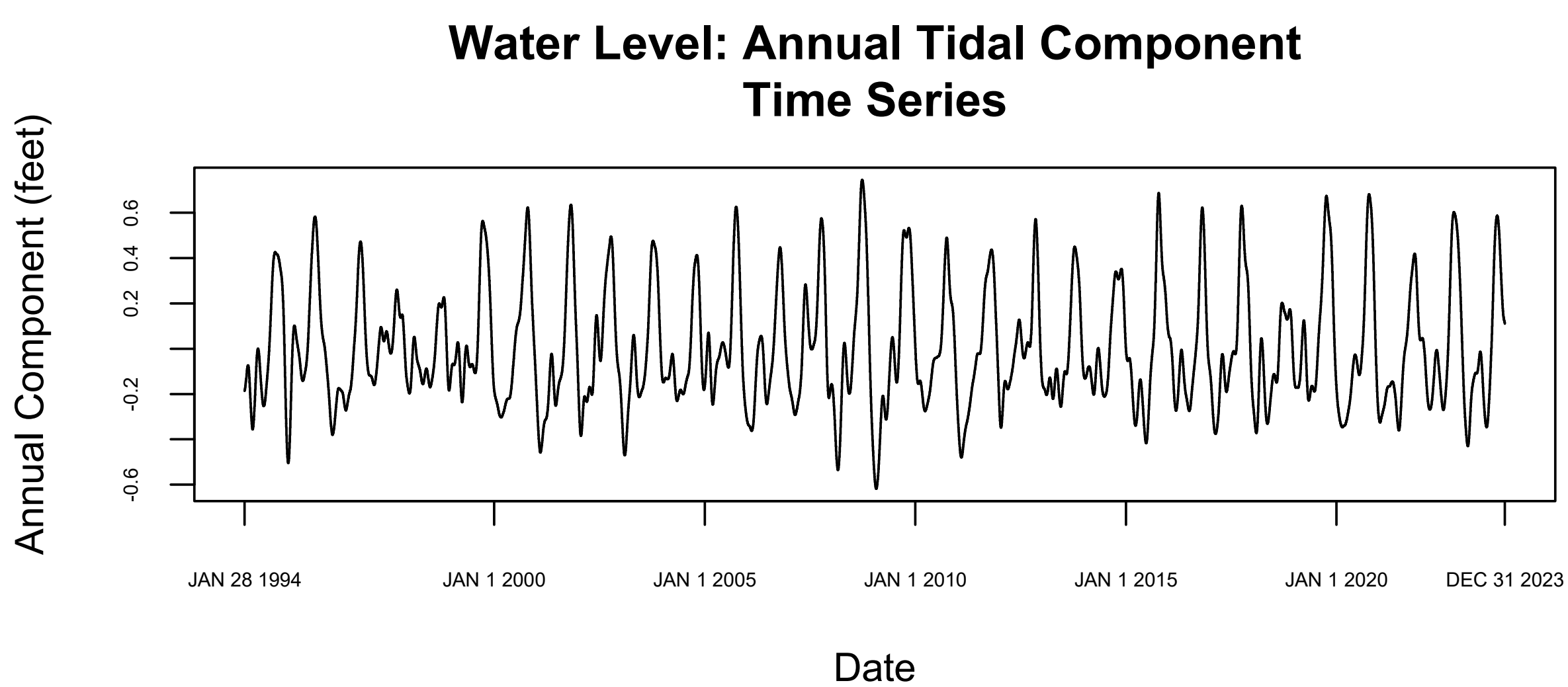

**FIGURE 3. Time Series* of the Annual Tidal Component for Hourly Water Level at Virginia Key, Florida from January 28, 1994, to December 31, 2024.**
*** Window width $m = 696$; number of iterations $k = 3$**

To do this, the average water level was computed for each hour of each date across the years from 1994 through 2023. As shown in Figure 4, average water levels indicated some periodicity over the course of a calendar year and this became more evident when smoothed using a KZ filter with a width of 29 days ($m$=696) with 3 iterations ($k$=3). The plot displays an "ascending" sinusoidal pattern with the end point of the average water level slightly higher than the starting point due to the long-term increase in water level as found in the analysis of long-term trends. It is also important to note that the highest crest in water level occurs just after the autumnal equinox (September 23), while the lowest trough in its greatest descent occurs just after winter solstice (December 21). The second highest crest occurs between vernal equinox (March 20) and summer solstice (June 23) while the lowest trough in its second greatest descent occurs approximately midway between summer solstice (June 21) and autumnal equinox (September 23).

Given these findings for average water level over the calendar year, the next step was to identify annual frequencies and their periods in the full 30-year time series dataset using one modification: the water level dataset was limited to weekly observations over this 30-year period to identify these frequencies more easily and was done by extracting every 168th observation (1 observation per week) from the hourly water level dataset. A KZ periodogram using a proportion of smoothing DZA = 0.01 identified two primary peaks: the first was the solar annual tide with a frequency of 0.019 (ln(amplitude$^2$ )=10.461, $CI_{.95}$=[9.155, 14.137] ) and a commensurate period of 368.42 days (approximately 1 calendar year) and the second was the solar semi-annual tide with a frequency of 0.038 (ln(amplitude$^2$)=9.115, $CI_{.95}$=[7.810, 12.792]) and a commensurate period of 184.21 days (approximately ½ calendar year). Given the arguments selected for the KZ periodogram with DZA smoothing, these frequencies and their respective periods are within the periodogram's limits of accuracy for the known annual and semi-annual tides as displayed in the filtered average tide heights for a calendar year shown in Figure 4.

### **3.4 Analysis of Daily Tides**

The next step was to extract the daily tidal component from the residual of the annual component. Residuals of the annual tidal component were determined by subtracting the annual component from the residuals of the long-term component and, again, these annual residuals centered around 0. The daily tidal component was extracted from these residuals using a KZ filter with 6-hour window ($m$=6) and 3 iterations ($k$=3). As noted above, there were two periods of missing observations; the first occurred in 1997, with 863 consecutive observations missing, and the second occurred in 2016, with 277 consecutive observations missing. Because the KZ filter window width was set at 6 and could accommodate only 18 consecutive missing observations, this and subsequent analyses were limited to the period of February 25, 2016, to December 31, 2023, with the daily tidal component displayed in Figure 5.

As in the KZ periodogram for the raw dataset from January 28, 1994, to December 31, 2023, the KZ periodogram for daily tidal component identified two sets of primary frequencies. In the first set, the strongest is the solar diurnal tide with a frequency of $\lambda = 0.042$ (ln(amplitude$^2$ )=8.862, $CI_{.95}$=[7.556, 12.538]) with a corresponding period of 23.81 hours, followed by the lunar diurnal tide with a frequency of $\lambda = 0.039$ (ln(amplitude$^2$ )=8.492, $CI_{.95}$=[7.187, 12.169]) with a corresponding period of 25.64 hours and, finally, the larger lunar elliptical diurnal tide with a frequency of $\lambda = 0.037$ (ln(amplitude$^2$ )=5.979, $CI_{.95}$=[4.674, 9.656] ) with a corresponding period of 27.03 hours. In the second set, the strongest is the principal lunar semi-diurnal tide with a frequency of $\lambda = 0.081$ (ln(amplitude$^2$ )=8.641, $CI_{.95}$=[7.336, 12.318]) with a corresponding period of 12.35 hours, followed by the principal solar diurnal tide with a frequency of $\lambda = 0.083$ (ln(amplitude$^2$ )=6.250, $CI_{.95}$=[4.944, 9.926]) with a corresponding period of 12.05 hours. Again, given the arguments selected for the KZ periodogram with DZA smoothing, these two sets of frequencies and respective periods are within the periodogram's limits of accuracy for the known annual and semi-annual tides present in the southwestern area of the North Atlantic Ocean.

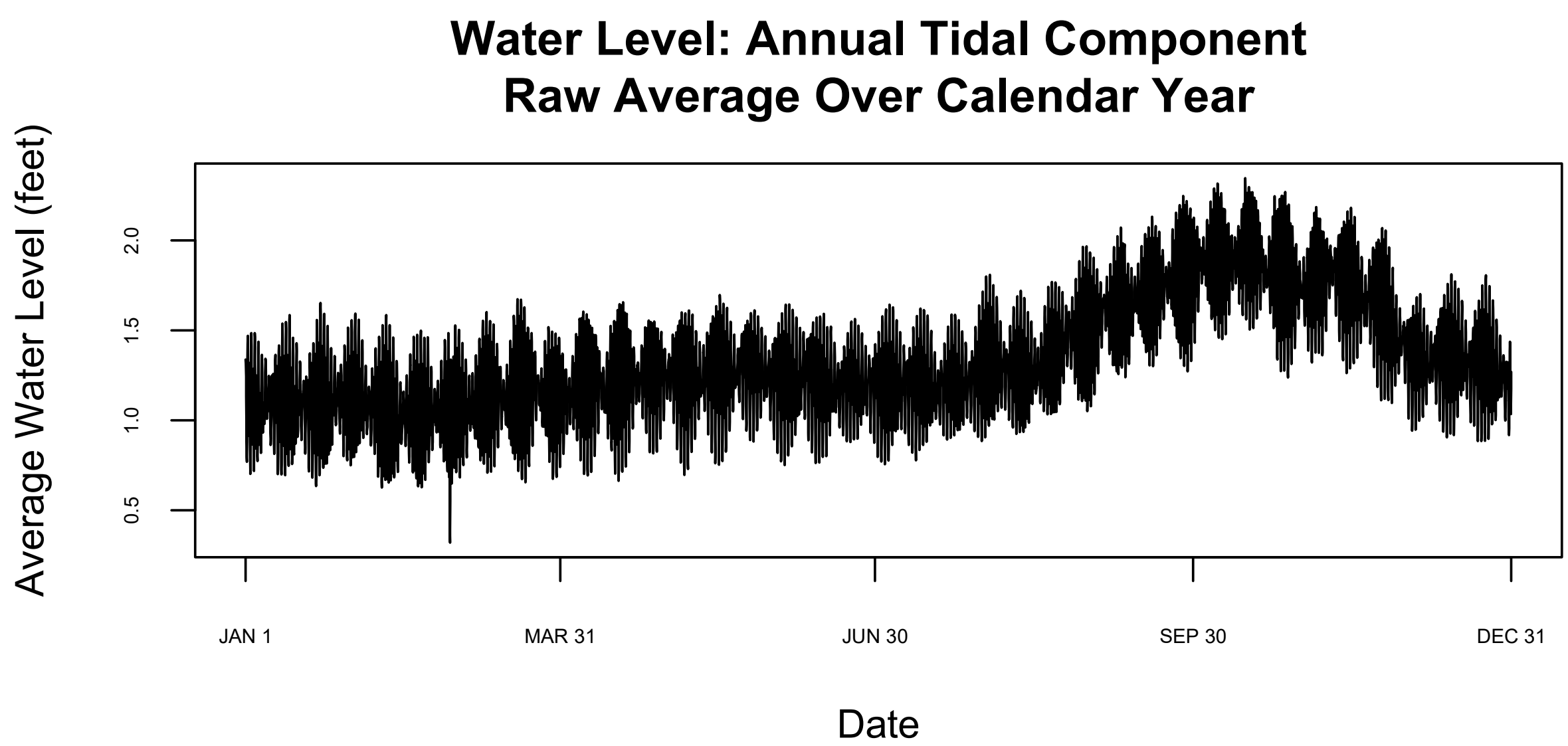

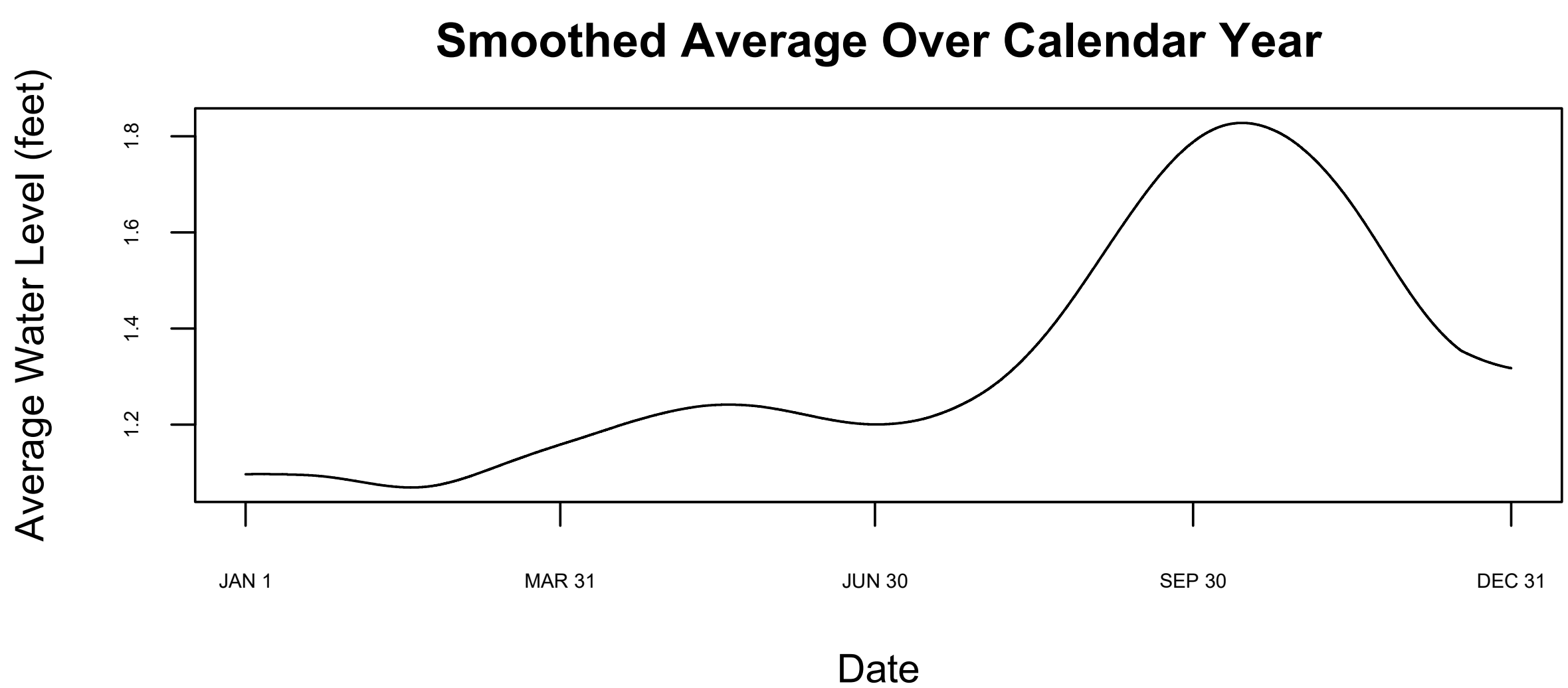

**FIGURE 4. Time Series* of Average Water Level over Calendar Year at Virginia Key, Florida, from January 1994, to December 2023.**
*** For smoothed time series: window width $m = 696$; number of iterations $k = 3$**

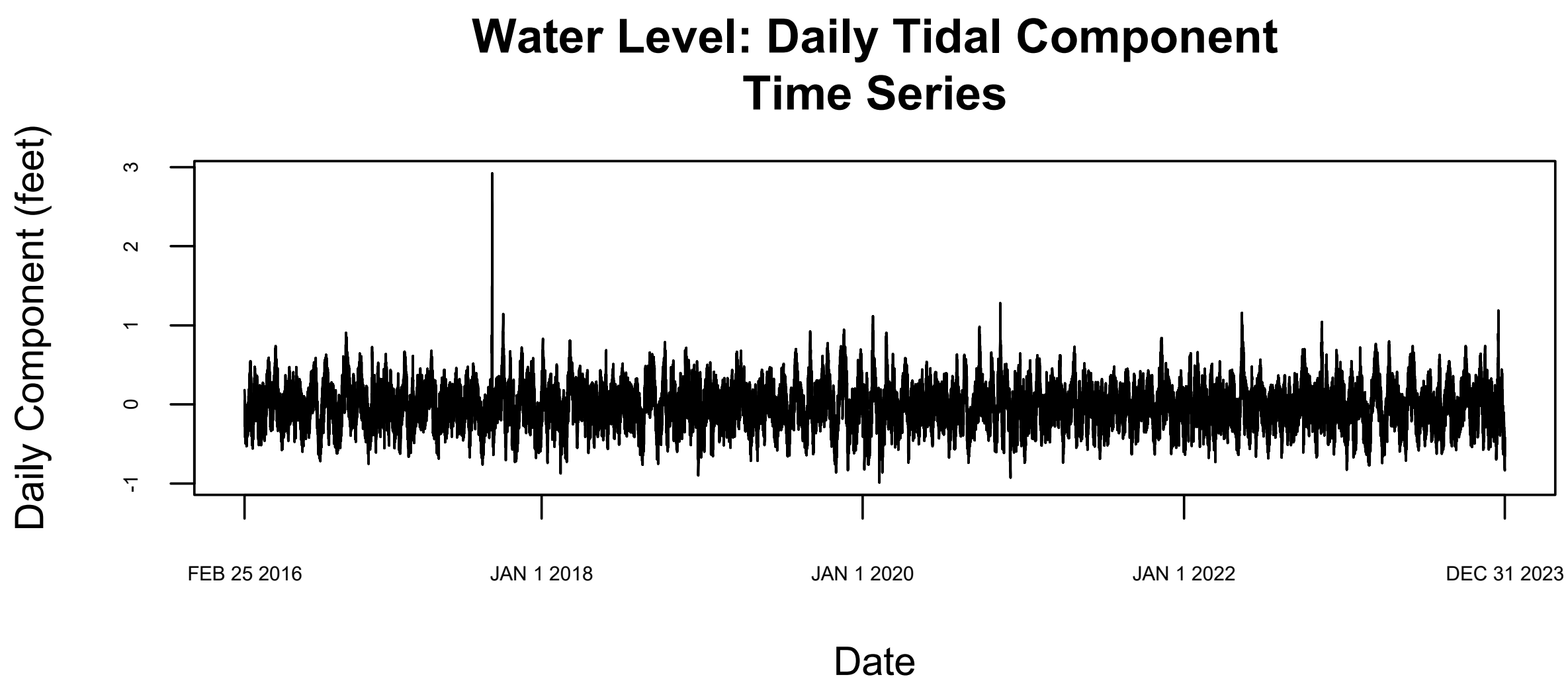

**FIGURE 5. Time Series of the Daily Tidal Component for Hourly Water Level at Virginia Key, Florida, from February 25, 2016, to December 31, 2024.**
*** Window width $m = 6$; number of iterations $k = 3$**

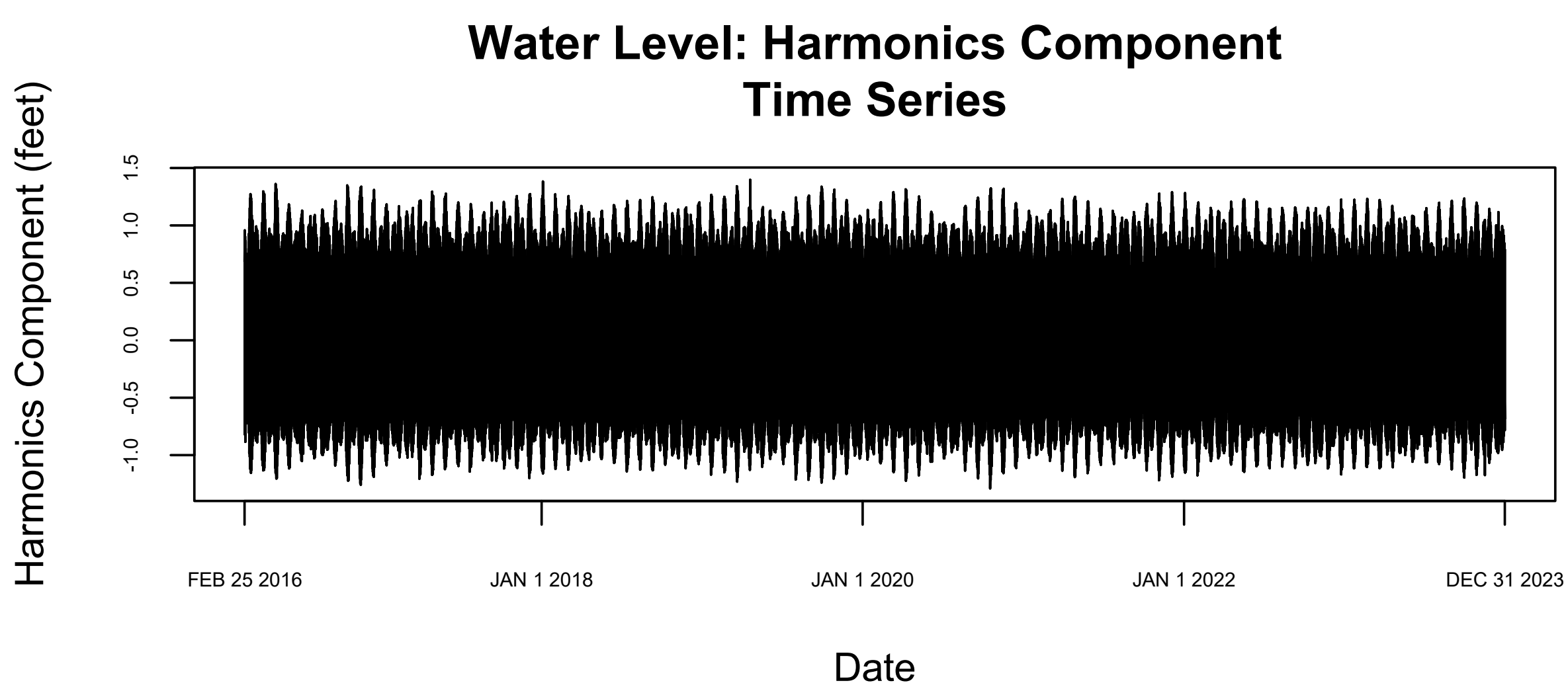

**FIGURE 6. Time Series of Higher Frequency Harmonics for Hourly Water Level at Virginia Key, Florida, from February 25, 2016, to December 31, 2023.**

## 3.5 Higher Frequency Harmonics

Finally, the higher frequency harmonics component (i.e., the residual of the daily tidal component) was extracted by subtracting the daily tidal component from the residual of the annual component, with the result displayed, above, in Figure 6. The multitude of patterns is indicative of multiple periodicities and is consistent with current tidal theory and research showing thousands of tidal harmonics (Casotto & Biscani, 2004; Doodson & Lamb, 1997).

## 3.6 Tidal and Harmonic Contributions to Total Water Level

As noted in the Methods section, time series data used in computation of tidal and higher frequency harmonic median, minima (2.5 percentile), and maxima (97.5 percentile) was limited to the period between February 2016 and December 2023. Overall, the median annual tidal contribution to water level was -0.0797 feet, with a minimum and a maximum of -0.358 feet and 0.619 feet, respectively. Similarly, the median daily tidal contribution to water level was -0.006 feet and the minimum and the maximum were found to be -0.499 feet and 0.529 feet, respectively. Finally, the median higher frequency harmonics contribution to water level was -0.016 feet and the minimum and the maximum were -0.950 feet and 1.015 feet, respectively. Combining the minima of annual tides, daily tides, and higher frequency harmonics indicates decreases as low as 1.807 feet below the water level determined by the long-term trend at a given time. Likewise, combining the maxima of annual tides, daily tides, and higher frequency harmonics indicates increases as high as 2.163 feet above the water level determined by the long-term trend at a given time.

## 3.7 Recapitulation

Table 1 summarizes key findings (i.e., frequencies, periods, frequency strength, strength confidence intervals) by tidal component for Virginia Key, Florida. As noted earlier, annual frequencies and their respective periods are consistent with known annual and semi-annual tides in this region, while daily frequencies and their respective periods are consistent with known diurnal and semi-diurnal tides in this region, all within the periodogram's limits for accuracy for the selected algorithm arguments. More importantly, Table 2 summarizes overall contributions to coastal water level by tidal component (annual, daily) and higher frequency harmonics, including their respective minima and maxima (i.e., 2.5 and 97.5 empirical percentiles). As noted earlier, the median contribution of both tidal components and of higher frequency harmonics approached 0 for each. However, by adding the sum of the maxima for annual tidal, daily tidal, and higher frequency harmonic components to the long-term water level, one can obtain a reasonable estimate of maximum total water levels. Similarly, by adding the sum of the minima of annual tidal, daily tidal, and higher frequency harmonic components to the long-term water level, one can obtain a reasonable estimate of minimum total water levels.

## 3.8 Predicted Water Levels

For the linear model, the long-term component was regressed on *time* for the period of February 25, 2016, to December 31, 2023, and the intercept was found to be .3953 ($p<0.0001$) with a regression coefficient for *time* of $5.294 \times 10^{-6}$ ($p<0.0001$). The intercept and coefficient were then used to estimate long-term water levels from January 1, 2024, to December 31, 2050; to ensure continuity in going from the observed long-term trend on December 31, 2023, to the predicted long-term trend on January 1, 2024, their difference was added to all predicted long-term water level values (i.e., 0.046 feet).

Observed long-term water levels from January 28, 1994, to December 31, 2023, were plotted, followed by predicted long-term water levels from January 1, 2024, to December 31, 2050. To determine minimum water levels over time, the sum of minima for annual and daily tides, and higher frequency harmonic components (i.e., -1.807 feet) was added to the respective observed and predicted long-term water levels. Likewise, to determine maximum water levels over time, the sum of maxima for annual and daily tides, and higher frequency harmonic components (i.e., 2.163 feet) was added to the respective observed and predicted long-term water levels.

For the quadratic model, the long-term component was regressed on *time* and $time^2$ for the period of February 25, 2016, to December 31, 2023, and the intercept was found to be 3.175 ($p<0.0001$) with a regression coefficient for *time* of $-1.929 \times 10^{-5}$ ($p<0.0001$) and a regression coefficient for $time^2$ of $5.393 \times 10^{-11}$ ($p<0.0001$). The intercept and coefficients were then used to estimate long-term water levels from January 1, 2024, to December 31, 2050; to ensure continuity in going from the observed long-term trend on December 31, 2023, to the predicted long-term trend on January 1, 2024, their difference was added to all predicted long-term water level values (i.e., 0.005 feet).

Observed long-term water levels from January 28, 1994, to December 31, 2023, were plotted, followed by predicted long-term water levels from January 1, 2024, to December 31, 2050. To determine minimum water levels over time, the sum of minima for annual and daily tides, and higher frequency harmonic components (i.e., -1.807 feet) was added to the respective observed and predicted long-term water levels. To determine the maximum water level over time, the sum of maxima for annual and daily tides, and higher frequency harmonic components (i.e., 2.163 feet) was added to the respective observed and predicted long-term water levels.

The results of these analyses are presented in Figure 7. According to the linear model, the total increase in long-term water levels between January 28, 1994, and December 31, 2050, is expected to be 2.02 feet, while according to the quadratic model, the total increase is expected to be 5.91 feet for the same time-period. In addition, the crests of annual and daily tides, along with higher frequency harmonics, will add up to 2.16 feet to the water level. Following the linear model, the lower bound of maximum total water level will be 4.18 feet greater than the median water level in 1994, while under the quadratic model, the upper bound of maximum total water level will be 8.07 feet greater than that median water level. Consequently, coastal water levels at Virginia Key, Florida, will be exceptionally high and any adverse weather events, such as hurricanes and tropical storms, could have even more devastating effects due to extensive flooding.

| Tide | $\lambda$ | Period | Frequency Strength Ln(Amplitude$^2$) | 95% CI Ln(Amplitude$^2$) |
|---|---|---|---|---|
| **ANNUAL AND SEMI-ANNUAL TIDES** | | | | |
| **Solar Annual** | 0.019 | 368.42 days | 10.461 | [9.155, 14.137] |
| **Solar Semi-Annual** | 0.038 | 184.21 days | 9.115 | [7.810, 12.792] |
| **DIURNAL AND SEMI-DIURNAL TIDES** | | | | |
| **Larger Lunar Elliptic Diurnal** | 0.037 | 27.03 hours | 5.979 | [4.674, 9.656] |
| **Lunar Diurnal** | 0.039 | 25.64 hours | 8.492 | [7.187, 12.169] |
| **Solar Diurnal** | 0.042 | 23.81 hours | 8.862 | [7.556, 12.538] |
| **Principal Lunar Semi-diurnal** | 0.081 | 12.35 hours | 8.641 | [7.336, 12.318] |
| **Principal Solar Semi-diurnal** | 0.083 | 12.05 hours | 6.250 | [4.944, 9.926] |

**TABLE 1. Summary of Tidal Frequency, Period, Log (Amplitude$^2$), and Log(Amplitude$^2$) 95% Confidence Interval by Tidal Component for Virginia Key, Florida.**

| Tidal Component | Median Contribution to Water Level | Minimum and Maximum of Component Contribution to Water Level |
|---|---|---|
| **Annual and Semi-Annual** | -0.080 | [-0.358, 0.619] |
| **Diurnal and Semi-Diurnal** | -0.006 | [-0.499, 0.529] |
| **Higher Frequency Harmonics** | -0.016 | [-0.950, 1.015] |

**TABLE 2. Summary Table of Median, Minimum, and Maximum Contribution to Coastal Water Level by Tidal Component for Virginia Key, Florida.**

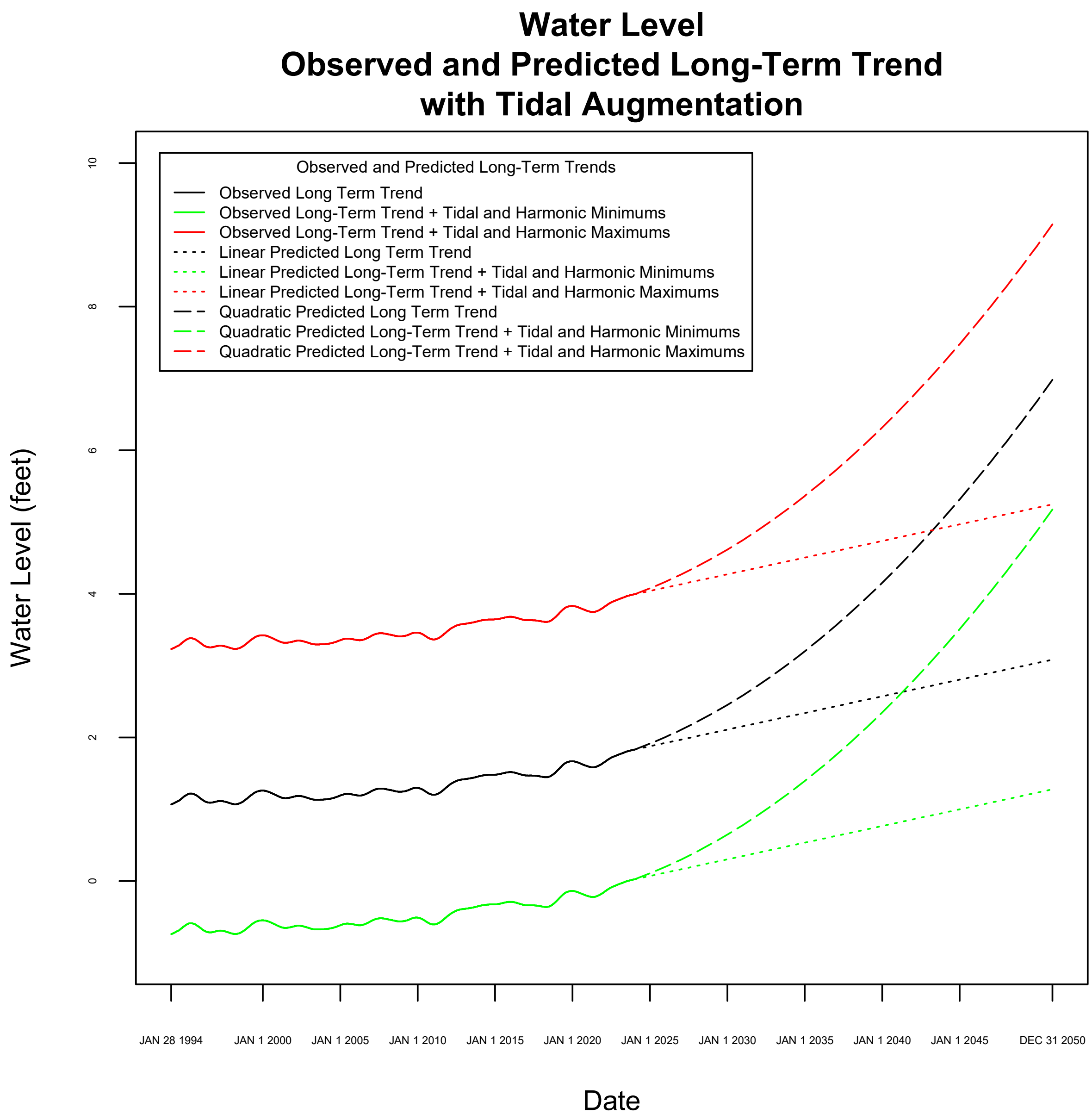

**FIGURE 7. Observed and Predicted Long-Term Trends in Water Level Augmented by Minima and Maxima of Annual Tides, Daily Tides, and Higher Frequency Harmonics. Note: 0 is a reference point equal to the average daily lowest low tide height from 1983 to 2001.**

## 4. Conclusion

Given the impact of global climate change on earth's ocean levels, this investigation substantiated the extent of this impact on coastal water levels for a specific location in the Southeastern United States. This section highlights the investigation's key findings, notes its limitations, outlines its implications for both emergency preparedness and proactive policies, and delineates recommendations for future research.

### 4.1 Key Findings

The long-term trend in coastal water level at Virginia Key, Florida, has resulted in an increase of .76 feet from January 1994 to December 2023, and is projected to increase an additional 1.26 feet by December 2050 according to a linear model and by an additional 5.15 feet according to a quadratic model. This increase is due to climate change – solar irradiation combined with air pollution due to human activity has increased atmospheric temperatures leading to melting of glaciers and the polar ice caps, with a corresponding increase in ocean water levels. While the combined tidal crests of annual tides, daily tides, and their higher frequency harmonics have been present throughout human history, the long-term increase in water levels caused by climate change places coastal residential communities, businesses, and wildlife habitats at considerable risk for flooding, particularly during catastrophic weather events such as hurricanes and tropical storms.

The previous work of Zurbenko, Potrzeba-Macrina, and Sun has modeled and predicted substantial changes for earth's climate, in general, and for its rising ocean levels, in particular. It is important to point out that these investigators made extensive use of Kolmogorov-Zurbenko analytic methods, including KZ filters and KZ periodograms. The current investigation built upon this work, in a practical way, to model and predict rising ocean levels for a specific location on the Southeastern coast of the United States. As such, it corroborates previous research because its substantive findings are in line with predictions of long-term increases in coastal water levels and demonstrates the utility of Kolmogorov-Zurbenko analytic methods through their use at a new site.

### 4.2 Limitations

There were three limitations in the current investigation. The first was missing data. Although KZ filters and KZ periodograms can effectively extract and identify signals even when there is an extensive amount of data that is missing completely at random (Iasonos & Zurbenko, 2002; Loneck et al., 2024; Zurbenko et al., 2020), analysis of an entire dataset is not possible when there is a large number of consecutive missing observations. This occurred twice in the current dataset (i.e., October 3, 1997, to November 8, 1997; February 14, 2016, to February 25, 2016). Nevertheless, a delimited dataset (i.e., February 25, 2016, to December 31, 2023) was used to extract and identify signals with shorter periods (i.e., diurnal and semi-diurnal tides, higher frequency harmonics) and the KZ filter and KZ periodogram were successful in doing so because the delimited dataset was of sufficient length.

The second limitation was the restricted range of smoothing windows available for KZ periodograms. Previous research has established the sensitivity, accuracy, and resolution of KZ periodograms with dynamic smoothing (e.g., DZA smoothing) in estimating signal frequencies (Loneck et al., 2024; Zurbenko et al., 2020); however, log-periodograms with static smoothing can be more precise in estimating signal amplitudes (Loneck et al., 2025). In the current investigation, tidal frequency estimates were within the limits of accuracy and resolution for the KZ periodograms, given the selected algorithm arguments, but precision of tidal and higher frequency harmonic amplitudes was limited because static smoothing windows are not yet available within the requisite software (Close et al., 2020).

The third limitation related to the processing capacity of available computer hardware. Identification of frequencies for signals with periods of 1 year (i.e., 8766 hours) requires KZ periodogram accuracy of at least 0.0001. This level of accuracy, in turn, requires that the KZ periodogram initial window width be set to $m_{Initial} = 10{,}000$. However, as $m_{Initial}$ increases, so does computing time and the time required for $m_{Initial} = 10{,}000$ surpassed what was reasonable for available computing equipment. Reanalyzing the data on a high-speed mainframe computer would not only enable identification of low frequency signals with corresponding long periods but would also improve the accuracy and resolution of all other identified frequencies.

## 4.3 Implications

Despite these limitations, the findings have practical short-term and long-term implications, while also serving as the basis for further investigations. In the short term, this investigation provides the foundation for more accurate prediction of flooding during severe weather events. As was done in this study, the observed long-term trend can be used to determine its contribution to total water level in the future for a given hour on a given day. Likewise, known periodicities of annual tides, daily tides, and their higher frequency harmonics can be used to determine water levels these components contribute to the total water level in the future for a given hour on a given day. If the approximate day and hour of a severe weather event are known, the preexisting total water level at event landfall can be predicted by summing the contributions of the long-term trend, annual tides, daily tides, and tidal harmonics. Consequently, emergency responders can be better prepared by knowing the extent of coastal flooding likely to occur.

In the long term, evidence of rising water levels indicates a need for policies and procedures that curtail human contributions to climate change, yet also address the upcoming long-term increases in solar irradiation. With regard to upcoming long-term increases in solar irradiation, such policies and procedures can include funds for relocation of residential communities and businesses away from coastal waterways and, to the extent possible, for protection of wildlife habitats. With regard to human influences, more stringent controls on pollutants that directly and indirectly lead to increases in atmospheric temperatures are required. While control of human influences will not eliminate the rise in ocean water levels, such controls can mitigate the severity of this rise in two ways. First, they will minimize the direct effect human factors have on climate change and, second, they will minimize any interactions between long-term increases in solar irradiation and human influences.

Finally, Kolmogorov-Zurbenko time series analytic methods were effective in extracting the long-term trend in water level attributed to climate change as well as in extracting and confirming known periodicities in water levels attributed to annual tides, daily tides, and their higher frequency harmonics. Consequently, future research in this area should also utilize these methods across a range of tidal classifications found in other coastal sites throughout the world (Webb, 2021). In addition to other coastal sites such as Miami that have semi-diurnal tides (i.e., two daily high tides and two daily low tides), other sites that have diurnal tides (i.e., one daily high tide and one daily low tide), and mixed semi-diurnal tides (i.e., two daily high tides and two daily low tides, with different respective heights) should also be studied. Investigations of long-term increases in water levels in concert with semi-diurnal tides could include other cities on the East Coast of North America, while long-term trends with diurnal tides can include cities on the Gulf of Mexico, the coast of Alaska, and Southeast Asia. Similarly, long-term trends along with mixed semi-diurnal tides can be studied on the Pacific Coast of North America. Through a complete study of all three tidal conditions throughout the world, a more comprehensive picture of the dynamic impact of climate change and tides on coastal water levels can be drawn.

## 5. References

Berg, R. (2016, January 29). The Alphabet Soup of Vertical Datums: Why MHHW is Mmm Mmm Good. *Inside the Eye*. https://noaanhc.wordpress.com/2016/01/29/the-alphabet-soup-of-vertical-datums-why-mhhw-is-mmm-mmm-good/

Cartwright, D. E. (2000). *Tides: A Scientific History*. Cambridge University Press.

Casotto, S., & Biscani, F. (2004). A fully analytical approach to the harmonic development of the tide-generating potential accounting for precession, nutation, and perturbations due to figure and planetary terms. *AAS/Division of Dynamical Astronomy Meeting #35*, *35*, 862.

Close, B., Zurbenko, I., & Sun, M. (2020). *kza: Kolmogorov-Zurbenko Adaptive Filters* (Version 4.1.0.1) [Computer software]. https://CRAN.R-project.org/package=kza

Darwin, G. H. (1997). XI. On the harmonic analysis of tidal observations of high and low water. *Proceedings of the Royal Society of London*, *48*(292–295), 278–340. https://doi.org/10.1098/rspl.1890.0041

DiRienzo, A. G., & Zurbenko, I. (1999). Semi-Adaptive Nonparametric Spectral Estimation. *Journal of Computational and Graphical Statistics*, *8*(1), 41–59. https://doi.org/10.1080/10618600.1999.10474800

Doodson, A. T., & Lamb, H. (1997). The harmonic development of the tide-generating potential. *Proceedings of the Royal Society of London. Series A, Containing Papers of a Mathematical and Physical Character*, *100*(704), 305–329. https://doi.org/10.1098/rspa.1921.0088

Federal Emergency Management Administration. (n.d.). *Community Report—Miami-Dade County, Florida | National Risk Index*. Retrieved July 15, 2024, from https://hazards.fema.gov/nri/report/viewer?dataLOD=Counties&dataIDs=C12086

Hyndman, R. J., & Fan, Y. (1996). Sample Quantiles in Statistical Packages. *The American Statistician*, *50*(4), 361–365. https://doi.org/10.1080/00031305.1996.10473566

Iasonos, A., & Zurbenko, I. (2002). The Effect of Missing Data on Linear Filters. *Proceedings of the American Statistical Association, Biometrics Section*, 1571–1576.

Kowalik, Z., & Luick, J. (2019). *Modern theory and practice of tide analysis and tidal power*. Austides Consulting.

Loneck, B., Zurbenko, I., & Valachovic, E. (2024). *Theoretical and Practical Limits of Kolmogorov-Zurbenko Periodograms with Dynamic Smoothing in Estimating Signal Frequencies* (arXiv:2007.03031v3). arXiv. https://doi.org/10.48550/arXiv.2007.03031

Loneck, B., Zurbenko, I., & Valachovic, E. (2025). *Theoretical and Practical Limits of Signal Strength Estimate Precision for Kolmogorov-Zurbenko Periodograms with Dynamic Smoothing* (arXiv:2412.07735v2). arXiv. https://doi.org/10.48550/arXiv.2412.07735

National Oceanic and Atmospheric Administration. (n.d.). *NOAA Tides & Currents: Tidal Datums*. Retrieved February 27, 2024, from https://tidesandcurrents.noaa.gov/datum_options.html

Potrzeba-Macrina, A., & Zurbenko, I. (2017). Computational Aspects of Spectral Estimations and Periodicities in Irregularly Observed Data. *Journal of Probability and Statistical Science*, *15*(2), 233–246.

Potrzeba-Macrina, A., & Zurbenko, I. (2019). Periods in Solar Activity. *Advances in Astrophysics*, *4*(2). https://doi.org/10.22606/adap.2019.42001

Potrzeba-Macrina, A., & Zurbenko, I. (2020). Numerical Predictions for Global Climate Changes. *World Scientific News*, *144*(2020), 208–225.

Potrzeba-Macrina, A., & Zurbenko, I. (2021). Multivariate Aspects of Global Warming. *World Scientific News*, *159*(2021), 81–94.

Rao, S. T., Zurbenko, I., Neagu, R., Porter, P. S., Ku, J. Y., & Henry, R. F. (1997). Space and Time Scales in Ambient Ozone Data. *Bulletin of the American Meteorological Society*, *78*(10), 2153–2166. https://doi.org/10.1175/1520-0477(1997)078<2153:SATSIA>2.0.CO;2

Valachovic, E. (2014). Solar Irradiation and the Annual Component of Skin Cancer Incidence. *Biometrics & Biostatistics International Journal*, *1*(3). https://doi.org/10.15406/bbij.2014.01.00017

Webb, P. (2021). *Introduction to Oceanography*. Roger Williams University. https://openlibrary-repo.ecampusontario.ca/jspui/handle/123456789/944

Yang, W., & Zurbenko, I. (2010). Kolmogorov–Zurbenko filters. *Wiley Interdisciplinary Reviews: Computational Statistics*, *2*(3), 340–351. https://doi.org/10.1002/wics.71

Zurbenko, I. (1986). *The Spectral Analysis of Time Series*. Elsevier North-Holland, Inc.

Zurbenko, I., & Potrzeba-Macrina, A. (2019a). Analysis of Regional Global Climate Changes due to Human Influences. *World Scientific News*, *132*, 1–15.

Zurbenko, I., & Potrzeba-Macrina, A. (2019b). Solar Energy Supply Fluctuations to Earth and Climate Effects. *World Scientific News*, *120*(2), 111–131.

Zurbenko, I., & Potrzeba-Macrina, A. (2021). Numerical Predictions for Rising Water Levels in the Oceans. *World Scientific News*, *152*(2021), 1–14.

Zurbenko, I., Smith, D., Potrzeba-Macrina, A., Loneck, B., Valachovic, E., & Sun, M. (2020). *High-Resolution Noisy Signal and Image Processing*. Cambridge Scholars Publishing.

Zurbenko, I., & Sun, M. (2015). Associations of Jet Streams with Tornado Outbreaks in the North America. *Atmospheric and Climate Sciences*, *5*(3), 336–344. https://doi.org/10.4236/acs.2015.53026

Zurbenko, I., & Sun, M. (2016). Jet Stream as a Major Factor of Tornados in USA. *Atmospheric and Climate Sciences*, *6*(2), 236–253. https://doi.org/10.4236/acs.2016.62020